\documentstyle[12pt]{article}
\input{epsf.tex}

\font\tbf = cmbx12

\begin{document}

  \begin{flushright} \begin{small}
  DF/IST-1.2001 \\gr-qc/0101052
  \end{small} \end{flushright}

\vskip 0.5cm

\begin{center}
{\tbf SCALAR, ELECTROMAGNETIC AND WEYL
PERTURBATONS OF BTZ BLACK HOLES: \\ QUASI NORMAL MODES} \\
\vskip 1cm
Vitor Cardoso\footnote{E-mail: vcardoso@fisica.ist.utl.pt} \\
\vskip 0.3cm
{\scriptsize  CENTRA, Departamento de F\'{\i}sica,
	      Instituto Superior T\'ecnico,} \\

{\scriptsize  Av. Rovisco Pais 1, 1096 Lisboa, Portugal,} \\

\vskip 0.6cm

Jos\'e P. S. Lemos\footnote{E-mail: lemos@kelvin.ist.utl.pt} \\
\vskip 0.3cm
{\scriptsize  CENTRA, Departamento de F\'{\i}sica,
	      Instituto Superior T\'ecnico,} \\
{\scriptsize  Av. Rovisco Pais 1, 1096 Lisboa, Portugal.}

\end{center} 
 
\bigskip

\begin{abstract}
\noindent

We calculate the quasinormal modes and associated frequencies of the
Ba\~nados, Zanelli and Teitelboim (BTZ) non-rotating black hole. This
black hole lives in 2+1-dimensions in an asymptotically anti-de Sitter
spacetime. We obtain exact results for the wavefunction and quasi normal
frequencies of scalar, electromagnetic and Weyl (neutrino) perturbations.

\strut  
\newline 
\end{abstract}

\newpage

\noindent
\section{ Introduction}
\vskip 3mm
When one is describing the evolution of some conservative
system, one often considers  a small perturbation
or a small departure from a known solution of the system,
and one generally arrives at a wave equation describing it.
For a system with no explicit time dependence, one finds the normal
mode solutions of the wave equation, satisfying certain boundary
conditions, and one can then specify completely the perturbation
as a linear superposition of these normal modes. In this case the
operator associated to the perturbation is self-adjoint, the frequencies are
real and the modes are complete.
 
However, when one deals with open dissipative systems, as it is the
case in this paper, such an expansion is not possible. Instead of
normal modes, one considers quasi normal modes (QNM) for which the
frequencies are no longer pure real, signalling that the system is
loosing energy. Although QNMs are in general not complete and
therefore insufficient to fully describe the dynamics (see
\cite{Beyer,Ching} and references therein), they nevertheless dominate
the signal during the intermediate stages of the perturbation, being
therefore extremely important.

QNMs of black holes were first numerically computed by
Chandrasekhar and Detweiler \cite{Chandra1}, and subsequent
numerical simulations \cite{Detweiler,Smarr,Anninos} 
showed that the amplitude is dominated, at intermediate
times, by a ringing signal due to the QNMs.  Aside from the pure
mathematical interest, black hole's QNM
calculations have been a very active field, and new methods, both
numerical and analytical have been developed (see 
\cite{kokkotas} for a review).

Up until very recently, all these works dealt with asymptotically
flat spacetimes. In the past few years there has been a growing interest
in asymptotically AdS (anti-de Sitter) spacetimes.
Indeed, the Ba\~nados-Teitelboim-Zanelli (BTZ) black hole in 2+1-dimensions 
\cite{Banados}, as well as black holes in 
3+1 dimensional AdS spacetimes with nontrivial topology 
(see, e.g. \cite{lemos1}), share with asymptotically flat spacetimes
the common property of both having well defined charges at infinity,
such as mass, angular momentum and electromagnetic charges, which makes them a
good testing ground when one wants to go beyond asymptotic flatness. 
Another very interesting aspect of these black hole solutions is related 
to the AdS/CFT (Conformal Field Theory) conjecture \cite{maldacena}. For instance, 
due to this AdS/CFT duality, quasi-normal frequencies 
in the BTZ black hole spacetime yield a prediction for the thermalization 
timescale in the dual two-dimensional CFT, which otherwise would be 
very difficult to compute directly. If one has, e.g., a 
10-dimensional type IIB supergravity, compactified into a
${\rm BTZ} \times
S^3 \times T^4$ spacetime, the scalar field used to perturb the BTZ black hole, 
can be seen as a type IIB dilaton which couples to a CFT field operator 
$\cal{O}$. Now, the BTZ in the bulk corresponds to to a thermal sate in 
the boundary CFT, and thus the bulk scalar perturbation corresponds to a thermal 
perturbation with nonzero $<\cal O>$ in the CFT.

There has been some recent work on perturbations of Schwarszchild AdS
spacetimes: Horowitz and Hubeny \cite{Horowitz} studied the QNM
frequencies for scalar perturbations in 4, 5 and 7 dimensions.  Wang
et al \cite{Wang1,Wang2} studied scalar perturbations and QNMs
on a Reissner-Nordstrom geometry, Chan and Mann \cite{chanmann} studied 
the QNM frequencies for a conformally coupled scalar field.
For work on BTZ black holes such as entropy of scalar fields, see \cite{Satoh}
and references therein.
  
In this paper we shall consider the QNMs of the
3D non-rotating BTZ black hole \cite{Banados}.
The non-rotating BTZ black hole metric for a spacetime with
negative cosmological constant, $\Lambda = -\frac{1}{l^2}$, is
given by
\begin{equation}
ds^{2}= (-M+\frac{r^2}{l^2})dt^{2}-
 (-M+\frac{r^2}{l^2})^{-1}dr^{2} - r^2d{\phi}^2                    
\,,
\label{2.1}
\end{equation}
where $M$ is the black hole mass. The horizon radius is given by $r_+
=M^{1/2}l$.  We shall in what follows suppose that the scalar,
electromagnetic and Weyl (neutrino) fields are a perturbation, i.e., they
propagate in a spacetime with a BTZ metric.  We will find that all
these fields obey a wave equation and the associated QNM 
are exactly
soluble yielding certain hypergeometric functions. As for the frequencies
one has exact and explicit results for scalar and electromagnetic
perturbations and numerical results for Weyl perturbations.
To our knowledge, this is the first exact solution of QNMs for 
a specific model (see \cite{Beyer}).

In section 2 we give the wave equation for scalar and electromagnetic 
perturbations, and find the QNMs themselves and their frequencies. 
In section 3 we find the wave equation for Dirac and Weyl (neutrino) 
perturbations and analyze their QNMs.


\noindent
\section{Perturbing a black hole with scalar and electromagnetic fields}


\subsection{The wave equation}
In this subsection we shall analyze the scalar and electromagnetic  
perturbations, which as we shall see yield the same effective potential, 
and thus the same wave equation.

First, for scalar perturbations, we are interested in solutions to 
the minimally coupled scalar wave equation 
\begin{equation}
{\Phi^\mu}_{;\,\mu}=0 \,, 
\label{minimalscalareq1}
\end{equation}
where, a comma stands for ordinary derivative and a semi-colon stands
for covariant derivative.
We make the following ansatz for the field $\Phi$
\begin{equation}
\Phi=\frac{1}{r^{1/2}}f(r)e^{-i\omega t}e^{im\phi}\,,
\label{ansatzforscalar}
\end{equation}
where $m$ is the angular quantum number. It is useful to use the tortoise coordinate  $r_*$
defined by the equation $dr_*=\frac{dr}{-M+\frac{r^2}{l^2}}$, and given implicitly by 
\begin{equation}
r=-M^{1/2}\coth(M^{1/2}r_*) \,,
\label{radiusastortoise}
\end{equation}
with $r_*\; \epsilon\;]-\infty,0]$, ($r_*=-\infty$ corresponds to $r=r_+$, and 
$r_*=0$ corresponds to $r=\infty$). 

With the ansatz (\ref{ansatzforscalar}) and the tortoise coordinate $r_*$, equation 
(\ref{minimalscalareq1}) is given by, 
\begin{equation}
\frac{d^2 f(r)}{d {r_*}^2} + (\omega - V(r))f(r)=0\,,
\label{minimalscalareq2}
\end{equation}
where,
\begin{equation}
V(r)=\frac{3r^2}{4 l^4} - \frac{M}{2 l^2}-\frac{M^2}{4 r^2}+\frac{m^2}{l^2}
 - \frac{Mm^2}{r^2}\,,
\label{potentialscalar}
\end{equation}
and it is implicit that $r=r(r_*)$. The rescaling to the the radial coordinate 
$\hat{r}=\frac{r}{l}$ and to the frequency $\hat{\omega}=\omega l$ is equivalent 
to take $l=1$ in (\ref{minimalscalareq2}) and (\ref{potentialscalar}), i.e., 
through this rescaling one measures the frequency and other quantities in 
terms of the AdS lengthscale $l$.

Now, the electromagnetic perturbations are governed by Maxwell's equations
\begin{equation}
{F^{\mu\nu}}_{\,;\nu}=0 \,\,, {\rm with} \,\, F_{\mu\nu}=A_{\nu,\mu}-
A_{\mu,\nu}\,,
\label{maxwellequation} 
\end{equation}
where $F_{\mu\nu}$ is the Maxwell tensor and $A_\mu$ is the electromagnetic 
potential. 
As the background is circularly symmetric, it would be advisable to expand $A_{\mu}$
in 3-dimensional vector spherical harmonics (see \cite{Edmonds} 
and \cite{Ruffini}):
\begin{eqnarray}
A_{\mu}(t,r,\phi)=
\left[ \begin{array}{c}g^{m}(t,r)\\h^{m}(t,r)
\\ k^{m}(t,r) \end{array}\right]e^{im\phi} \,, 
\label{empotentialdecomposition}
\end{eqnarray}
where $m$ is again our angular quantum number, and 
this decomposition is similar to the one in eigenfunctions
of the total angular momentum in flat space \cite{Edmonds}. 

However, going through the same steps one finds that the equation 
for electromagnetic perturbations is the same as the one for scalars, 
equation (\ref{minimalscalareq2}). The reason is that in three dimensions 
the 2-form Maxwell field $F=F_{\mu\nu}dx^\mu\wedge dx^\nu$ is dual to 
a 1-form $d\Phi$.

\subsection{QNMs for scalar and electromagnetic perturbations}

Although a precise mathematical definition for a QNM can be given, as a pole
in the Green's function \cite{kokkotas}, we shall follow a more
phenomenological point of view. A QNM  describes the
decay of the field in question.  For the equation
(\ref{minimalscalareq2})  
it is defined as a corresponding solution 
which (i) near the horizon is purely ingoing, $\sim
e^{i\omega r_*}$, corresponding to the existence of a black hole,
 and (ii) 
near infinity is 
purely outgoing, $\sim e^{-i\omega r_*}$, (no
initial incoming wave from infinity is allowed). One can see that the potential $V(r)$ 
diverges at infinity, so we require that the perturbation
vanishes there (note that $r=\infty$ corresponds to a finite value of $r_*$, namely
$r_*=0$). This vanishing of the solution at $\infty$ will only be possible for a 
discrete set of complex frequencies $\omega$ called quasinormal frequencies.

 \subsubsection{Exact calculation}

Puting $l=1$  and 
using the coordinate $r_*$, the wave equation (\ref{minimalscalareq2}) takes the form
\begin{eqnarray}
&\frac{\partial^{2} a(r)}{\partial r_*^{2}} +
&\nonumber\\&
\left[\omega^2
 -\frac{3M}{4\sinh(M^{1/2}r_*)^2}+\frac{M}{4\cosh(M^{1/2}r_*)^2}+
\frac{L^2}{\cosh(M^{1/2}r_*)^2}\right]a(r)=0 \,.&
\label{scalarmaxwellequation2}
\end{eqnarray}
On going to a new variable $x=\frac{1}{\cosh(M^{1/2}r_*)^2}$,
$x\; \epsilon\;[0,1]$ equation (\ref{scalarmaxwellequation2}) 
can also be written as 
\begin{equation}
4x(1-x)\partial_x^2a +(4-6x)\partial_x\psi +\bar{V}(x)a=0 \,,
\label{scalarmaxwellequation3}
\end{equation}
where 
\begin{equation}
\bar{V}(x)=\frac{1}{4x(1-x)}\left[\frac{4\omega^2(1-x)}{M}-3x
-x(1-x)-\frac{4m^2x(1-x)}{M}\right]\,.
\end{equation}
By changing to a new wavefunction y (see \cite{Uvarov} for details), through
\begin{equation}
\psi \rightarrow \frac{(x-1)^{3/4}}{x^{\frac{i\omega}{2M^{1/2}}}} y \,,
\label{step}
\end{equation}
equation (\ref{scalarmaxwellequation3})
can be put in the canonical form \cite{Uvarov,Stegun}
\begin{equation}
x(1-x)y'' + [c-(a+b+1)x]y' -aby =0 \,,
\label{scalarmaxwellequation4}
\end{equation}
with 
$a=1+\frac{im}{2M^{1/2}}-\frac{i\omega}{2M^{1/2}} $, 
$b=1-\frac{im}{2M^{1/2}}-\frac{i\omega}{2M^{1/2}} $, 
and $c=1-\frac{i\omega}{M^{1/2}}$,
which is a
standard hypergeometric equation. The hypergeometric equation has
three regular singular points at $x=0,x=1,x=\infty$, and has two
independent solutions in the neighbourhood of each singular point. We
are interested in solutions of (\ref{scalarmaxwellequation4}) in the range [0,1],
satisfying the boundary conditions of ingoing waves near $x=0$, and
zero at $x=1$. One solution may be taken to be
\begin{equation}
y=(1-x)^{c-a-b}F(c-a,c-b,c;x) \,,
\label{hypersolution}
\end{equation}
where $F$ is a standard hypergeometric function of the second kind.
Imposing $y=0$ at $x=1$, and recalling that 
$F(a,b,c,1)=\frac{\Gamma(c)\Gamma(c-a-b)}{\Gamma(c-a)\Gamma(c-b)}$, we get
\begin{equation}
a=-n \,, {\rm or}\;  b=-n \,,
\label{parameters}
\end{equation}
with $n=0,1,2,...\,$, so that the quasi normal frequencies are given by
\begin{equation}
\omega=\pm m -2iM^{1/2}(n+1).
\label{frequency1}
\end{equation}
The lowest frequencies, namely those with $n=0$ and $m=0$ had already been
obtained by \cite{Govinda} and agree with our results. 

\subsubsection{Numerical calculation of the frequencies}

In order to check our results, we have also computed numerically the
frequencies. By going to a
new variable $ z=\frac{1}{r} $, $h=\frac{1}{r_+}$ one can put the wave
equation (\ref{minimalscalareq2}) in the form (see
\cite{Horowitz} for further details)
\begin{equation}
s(z)\frac{d^2}{dz^2}\Theta +t(z)\frac{d}{dz}\Theta+u(z)\Theta=0 \,, 
\label{scalarmaxwellequation5}
\end{equation}
where $\Theta=e^{i\omega r_*} a(r)$, $s(z)=z^2-Mz^4$, $t(z)=2Mz^3-2i \omega z^2$ 
and $u(z)=\frac{V}{-M+\frac{1}{z^2}}$, with $V$ given by 
(\ref{potentialscalar}). 
Now, $z$ $\epsilon$ $[0,h]$ and one sees that in this range, the differential
equation has only regular singularities at $z=0$ and $z=h$, so it has by Fuchs theorem
a polynomial solution. We can now use Fr\"{o}benius method (see for example \cite{Arfken})
and look for a solution of the form $\Theta(z)= \sum_{n=0}^{\infty} \theta_{n(\omega)} (z-h)^n (z-h)^{\alpha} $,
where $\alpha$ is to be taken from the boundary conditions. Using the boundary condition of only ingoing waves 
at the horizon, one sees \cite{Horowitz} that $\alpha=0$. So the final outcome is that $\Theta\,$ can be expanded as 
\begin{eqnarray}
\Theta(z)= \sum_{n=0}^{\infty} \theta_{n(\omega)} (z-h)^n \,.
\label{expansion1} 
\end{eqnarray}
Imposing  now the second boundary condition, $\Theta=0$ at infinity ($z=0$) one gets
\begin{equation}
\sum_{n=0}^{\infty} \theta_{n(\omega)}(-h)^n=0 \,.
\label{expansion2} 
\end{equation}
The problem is reduced to that of finding a numerical solution of the
polynomial equation (\ref{expansion2}).  The numerical roots for
$\omega$ of equation (\ref{expansion2}) can be evaluated resorting to
numerical computation. Obviously, one cannot determine the full sum in
expression (\ref{expansion2}), so we have to determine a partial sum
from 0 to N, say, and find the roots $\omega$ of the resulting
polynomial expression.  We then move onto the next term N+1 and
determine the roots.  If the method is reliable, the roots should
converge. We have stopped our search when a 3 decimal digit precision 
was achieved.
We have computed the lowest frequencies for some parameters of the
angular quantum number $m$ and horizon radius $r_+$.  The frequency is written as
$\omega = \omega_r + i\omega_i$, where $\omega_r$ is the real part of
the frequency and $\omega_i$ is its imaginary part.

In tables 1 and 2 we list the numerical values of the lowest
QNM frequencies, for $m=0$ and $m=1$, respectively, and 
for selected values of the black hole mass.
\begin{center}
\begin{tabular}{|l|l|l|l|l|}  \hline 
\multicolumn{5}{|c|}{$m=0$} \\  \hline 
\multicolumn{1}{|c|}{} &
\multicolumn{2}{c|}{ Numerical} &
\multicolumn{2}{c|}{ Exact} \\ \hline
$M^{1/2}$   &  $\omega_r$ &  $-\omega_i$ &  $\omega_r$  & $-\omega_i$ \\ \hline
$\frac{1}{2}$   & 0.000 & 1.000 & 0 & 1  \\ \hline
 1   & 0.000 & 2.000 & 0 & 2  \\ \hline
 5    & 0.000 & 10.000 & 0 & 10  \\ \hline
 10   &  0.000 & 20.000  & 0 & 20  \\ \hline
 50   & 0.000 & 100.000 & 0 & 100  \\ \hline
 100   & 0.000  & 200.000 & 0 & 200   \\ \hline
 1000   & 0.000  & 2000.000 & 0& 2000   \\ \hline
\end{tabular}
\end{center}
\vskip 1mm
\centerline{Table 1. Lowest ($n=0$) QNM frequencies for $m=0$. }

\vskip 1cm

\begin{center}
\begin{tabular}{|l|l|l|l|l|}  \hline 
\multicolumn{5}{|c|}{$m=1$} \\  \hline 
\multicolumn{1}{|c|}{} &
\multicolumn{2}{c|}{ Numerical} &
\multicolumn{2}{c|}{ Exact} \\ \hline
$M^{1/2}$   &  $\omega_r$ &  $-\omega_i$ &  $\omega_r$  & $-\omega_i$ \\ \hline
$ \frac{1}{2} $  & 1.000 & 1.000 & 1 & 1  \\ \hline
 1   & 1.000 & 2.000 & 1 & 2  \\ \hline
 5    & 1.000 & 10.000 & 1 & 10  \\ \hline
 10   &  1.000 & $20.000$  & 1 & 20  \\ \hline
 50   & 1.000 & 100.000 & 1 & 100  \\ \hline
 100   & 1.000  & 200.000 & 1 & 200   \\ \hline
 1000   & 1.000  & 2000.000 & 1& 2000   \\ \hline
\end{tabular}
\end{center}
\vskip 1mm
\centerline{ Table 2. Lowest ($n=0$) QNM frequencies for $m=1$. }
\vskip 1cm

The numerical results agree perfectly with (\ref{frequency1}), and
one sees that the imaginary part of the frequency scales with the 
horizon whereas the real part depends only on the angular index $m$.

\section{Perturbing a black hole with Dirac and \hfill\break Weyl spinor fields}

\subsection{The wave equation}


We shall develop Dirac's equation for a massive spinor, and then
specialize to the massless case.
The two component massive spinor field $\Psi$, with mass
$\mu_s$ obeys the covariant
Dirac equation
\begin{equation}
i\gamma^\mu \nabla_{\mu}\Psi -\mu_s \Psi=0\, ,
\label{diracequation}
\end{equation}
where $\nabla_{\mu}$ is the spinor covariant derivative defined by
$\nabla_{\mu} = \partial_{\mu}+\frac{1}{4} \omega^{ab}_{\mu} \gamma _{[a} 
\gamma_{b]}$, and $\omega^{ab}_{\mu}$ is the spin connection, which 
may be given in terms of the tryad $e_a^\mu$. 

As is well known there are two inequivalent two dimensional irreducible
representations of the $\gamma$ matrices in three spacetime dimensions.
The first may be taken to be $\gamma^0=i\sigma^2\,,\gamma^1=\sigma^1$, and 
$\gamma^2=\sigma^3$, where the matrices $\sigma^k$ are the Pauli matrices.
The second representation is given in terms of the first by a minus 
sign in front of the Pauli matrices. From equation (\ref{diracequation}), 
one sees that a Dirac particle with mass $\mu_s$ in the first representation 
is equivalent to a Dirac particle with mass $-\mu_s$ in the second 
representation. To be definitive, we will use the first representation, but the 
results can be interchanged to the second one, by substituting $\mu_s\rightarrow-\mu_s$.
For Weyl particles, $\mu_s=0$, both representations yield the same results.

Again, one can separate variables by setting
\begin{eqnarray}
\Psi(t,r,\phi)=
\left[ \begin{array}{c}\Psi_1(t,r)\\ \Psi_2(t,r)
\end{array}\right]e^{-i\omega t}e^{im\phi} \,.
\label{diracdecomposition}
\end{eqnarray}
On substituting this decomposition into Dirac's equation (\ref{diracequation}) 
we obtain
\begin{eqnarray}
-\frac{i(M-2r^2)}{2\Delta^{1/2}}r\Psi_2 +i\Delta^{1/2}\partial _r \Psi_2
+\frac{r^2\omega}{\Delta^{1/2}}\Psi_2=(m+\mu_s)\Psi_1 \,,
\label{diracequation2a}
\\
-\frac{i(M-2r^2)}{2\Delta^{1/2}}r\Psi_1 +i\Delta^{1/2}\partial _r \Psi_1
+\frac{r^2\omega}{\Delta^{1/2}}\Psi_1=(m+\mu_s)\Psi_2 \,,
\label{diracequation2b}
\end{eqnarray}
where we have put $\Delta=-Mr^2+\frac{r^4}{l^2}$, we have restored the 
AdS lengthscale $l$, and in general we follow
Chandrasekhar's notation \cite{Chandra2}. Defining $R_1$, $R_2$, and $\hat{m}$
through the relations
\begin{eqnarray}
\Psi_1=i\Delta^{-1/4}R_1\,, 
\label{transform1}\\
\Psi_2= \Delta^{-1/4}R_2 \,,
\label{transform2}\\  
m=i\hat{m}\,,
\label{transform3}
\end{eqnarray} 
we obtain,
\begin{eqnarray}
(\partial_{r_*}-i\omega)R_2=\frac{i\Delta^{1/2}}{r^2}(\hat{m}-i\mu_sr)R_1 \,,
\label{diracequation3a}\\
(\partial_{r_*}+i\omega)R_1=\frac{i\Delta^{1/2}}{r^2}(\hat{m}+i\mu_sr)R_2 \,.
\label{diracequation3b}
\end{eqnarray}
Defining now $\nu$, $\Upsilon_1$ $\Upsilon_2$, and $\hat{r}_*$ through the 
relations
\begin{eqnarray}
\nu=\arctan(\frac{\mu_sr}{\hat{m}})\,,
\label{definition1a}\\
R_1=e^{\frac{i\nu}{2}}\Upsilon_1 \,,
\label{definition1b}\\
R_2=e^{\frac{-i\nu}{2}}\Upsilon_2 \,,
\label{definition1c}\\
\hat{r}_*=r_* +\frac{1}{2\omega}\arctan(\frac{\mu_sr}{\hat{m}})\,, 
\label{definition1d}
\end{eqnarray}
we get
\begin{eqnarray}
(\partial_{\hat{r}_*}-i\omega)\Upsilon_2=W\Upsilon_1 \,,
\label{diracequation4a}\\
(\partial_{\hat{r}_*}-i\omega)\Upsilon_2=W\Upsilon_2 \,,
\label{diracequation4b}
\end{eqnarray}
where,
\begin{eqnarray}
W=\frac{i\Delta^{1/2}(\hat{m}^2+\mu_s^2r^2)^{3/2}}{r^2(\hat{m}^2+\mu_s^2r^2)
+\frac{\hat{m}\mu_s\Delta}{2\omega}} \,. 
\label{potentialdirac1}
\end{eqnarray}
Finally, putting $Z_{\pm}=\Upsilon_1 \pm \Upsilon_2$ we have
\begin{eqnarray}
(\partial_{\hat{r}_*}^2+\omega^2)Z_{\pm}=V_{\pm}Z_{\pm} \,,
\label{diracequation5}
\end{eqnarray}
with
\begin{equation}
V_{\pm}=W^2 \pm \frac{dW}{d\hat{r}_*} \,.
\label{potentialdirac2}
\end{equation}
We shall be concerned
with massless spinors ($\mu_s=0$) for which $ \hat{r}_*=r_*$, and 
$W=\frac{i\Delta^{1/2}\hat{m}}{r^2}$. Thus, 
\begin{eqnarray}
V_{\pm}=\frac{m^2}{r^2}(\frac{r^2}{l^2}-M) \pm \frac{Mm}{r^2}
(\frac{r^2}{l^2}-M)^{1/2} \,.
\label{potentialdirac3}
\end{eqnarray}
In the form (\ref{potentialdirac2}) one immediatly recognizes that the two 
potentials $V_+$ and $V_-$ should yield the same spectrum. In fact
they are, in SUSY language, superpartner potentials derived
from a superpotential W (see \cite{Cooper}).
Once again, we can rescale $r$ and take $l=1$, by measuring
everything in terms of l.

\subsection{ QNMs for Weyl perturbations}
Similarly, the wave equation (\ref{diracequation5}) for Weyl 
(until recently also called neutrino) perturbations
may be put in the form
\begin{equation}
\partial_{r_*}^2 Z_{\pm}+\left[\omega^2-
m\left(\frac{m}{\cosh(M^{1/2}r_*)^2} \pm M^{1/2}\frac{\sinh(M^{1/2}r_*)^2}
{\cosh(M^{1/2}r_*)^2}\right)\right]Z_{\pm}=0 \,.
\label{neutrinoequation1}
\end{equation} 
Going to a new independent variable, $x=-\sinh(M^{1/2}r_*)$,
$x$ $\epsilon$ $[\infty,0]$, we can write
\begin{equation}
(1+x^2)Z'' +xZ' +[\frac{\frac{\omega^2(1+x^2)}{M}-\frac{m^2}{M}
\pm\frac{mx}{M^{1/2}}}{1+x^2}]Z=0\,. 
\label{neutrinoequation2}
\end{equation}
By changing the wavefunction Z to $\chi$
\begin{equation}
\chi= e^{(\frac{M^{1/2}x-2m}{2M^{1/2}}-\frac{x}{2})\arctan(x)}\,,
\label{change}
\end{equation}
we have 
\begin{equation}
(1+x^2)\chi'' +(\frac{2m}{M^{1/2}}+x)\chi' +(\frac{\omega^2}{M})\chi=0 \,.
\label{neutrinoequation3}
\end{equation}
On putting $s=\frac{1+iz}{2}$, $s$ $\epsilon$ $[\frac{1}{2},i\infty]$, 
we have again the hypergeometric equation (\ref{scalarmaxwellequation4}), 
with $a=\frac{i\omega}{M^{1/2}}$, $b=-\frac{i\omega}{M^{1/2}} $, 
and $c=\frac{1}{2}\pm\frac{im}{M^{1/2}}$,
so that
the solution to the wave equation is again specified around each
singular point, and is given by the analytic continuation of the
standard hypergeometric function to the complex plane \cite{Uvarov,Stegun}.

Since infinity is located at $s=\frac{1}{2}$, there is no easy
way to determine the QNM frequencies, so we have to resort to
numerical calculations. If we put (\ref{potentialdirac1}) in the form
(\ref{scalarmaxwellequation5}) one again sees that it has no essential singularities,
so the numerical method just outlined in the previous section may be
applied. Moreover, since $V_+$ and $V_-$ have the same spectrum
\cite{Cooper} and the same QNM frequencies \cite{Chandra2} we
need only to workout the frequencies for one of them. In table 3
we present the numerical results for the QNM frequencies for 
neutrino perturbations and for selected values of the black hole mass.

\begin{center}
\begin{tabular}{|l|l|l|l|l|}  \hline 
\multicolumn{3}{|c|}{$m=1$} \\  \hline 
\multicolumn{1}{|c|}{} &
\multicolumn{2}{c|}{ Numerical} \\ \hline
$M^{1/2}$   &  $\omega_r$ &  $-\omega_i$ \\ \hline
 2   & 0.378 & 2.174   \\ \hline
 5   & 0.316 & 5.027    \\ \hline
 10    & 0.224 & 10.006   \\ \hline
 50   &  0.099 & 50.001    \\ \hline
 100   & 0.071 & 100.000  \\ \hline
 500   & 0.0316  & 500.000    \\ \hline
\end{tabular}
\end{center}
\vskip 1mm
\centerline{Table 3. Lowest QNM frequencies for $m=1$. }
\vskip 1mm

For large black holes one can see that the imaginary part of the 
frequencies scale with the horizon ($r_+=M^{1/2}$),
just as in the scalar and electromagnetic case. We have also computed
some higher modes, and the real part of the frequency $\omega_r$,  does not seem
to depend on which mode we are dealing with, just as in the scalar and
electromagnetic case.

\section{Conclusions}
 We have computed the scalar,  electromagnetic and
neutrino QNM of BTZ black holes.
These modes dictate the late time behaviour of the fields.
In all cases, these modes scale with the horizon radius,
at least for large black holes and, since the decay of the perturbation
has a timescale $\tau = \frac{1}{\omega_i}$, this means
that the greater the mass, the less time it takes to approach
equilibrium. We have also found that for large black holes, 
the QNM frequencies are proportional to the black hole 
radius. Since the temperature of a BTZ black is proportional 
to the black hole radius, the QNM frequencies scale with 
the temperature, as a simple argument indicates \cite{Horowitz}.
For the study of QNM of 3+1-dimensional spherical, as well as 
the toroidal black holes found by Lemos \cite{lemos2} see \cite{Cardoso}.

Is there, for small black holes,
any relation with these quasinormal modes and critical
phenomena as speculated by Horowitz and Hubeny \cite{Horowitz}? 
Though that would be an extremely interesting relation, we still
are not able to answer that.

\vskip 1cm

\section*{Acknowledgments} 
This work was partially funded by FCT through project Sapiens.
One of us (V. C.) acknowledges finantial support 
from the portuguese FCT through PRAXIS XXI
programme. 

\vskip 1cm


\end{document}